\documentclass[conference]{IEEEtran}
\usepackage{fancyhdr}

\IEEEoverridecommandlockouts

\usepackage{dingbat,utfsym}
\usepackage{cite}
\usepackage[hyphens]{url}
\usepackage{tikz}
\usepackage{braket}
\usepackage{subfig}
\usepackage{xcolor}
\usepackage{amsmath}
\usepackage{amssymb}
\usepackage{balance}
\usepackage{environ}
\usepackage{physics}
\usepackage{colortbl}
\usepackage{enumitem}
\usepackage{amsfonts}
\usepackage{booktabs}
\usepackage{graphicx}
\usepackage{textcomp}
\usepackage{algorithm}
\usepackage{algpseudocode}
\usepackage[normalem]{ulem}
\usepackage[most]{tcolorbox}
\usepackage{fontawesome5} % for gear and checkmark symbols
\usepackage{pdfpages}

\def\BibTeX{{\rm B\kern-.05em{\sc i\kern-.025em b}\kern-.08em
    T\kern-.1667em\lower.7ex\hbox{E}\kern-.125emX}}

\definecolor{red}{rgb}{1.0, 0.0, 0.0}
\definecolor{blue}{rgb}{0.0, 0.0, 1.0}
\definecolor{green}{rgb}{0.0, 0.62, 0.0}

\newcommand{\xmark}{\textcolor{red}{\raisebox{0.2ex}{\usym{1F5F4}}}}
\newcommand{\cmark}{\textcolor{green}{\checkmark{}}}

\newcommand{\sol}{\textsc{Parallax}}
\newcommand{\geyser}{\textsc{Geyser}}
\newcommand{\graphine}{\textsc{Graphine}}
\newcommand{\eldi}{\textsc{Eldi}}
\newcommand{\olsq}{\textsc{DPQA}}

\newcommand{\solution}{\faIcon{check-square}}

\newtcolorbox{challengebox}[1]{
    colback=orange!5!white,
    colframe=orange!75!black,
    coltitle=white,
    colbacktitle=white!75!black, % Background color of title
    title={\strut#1}, % Strut for consistent height
    enhanced,
    attach boxed title to top left={yshift=-2mm, xshift=2mm},
    boxed title style={colframe=orange!75!black, colback=white}, % Style for title background
    left=2mm,
    right=2mm,
    fonttitle=\bfseries,
    halign title=flush center,
}

\newtcolorbox{hardwarechallengebox}[1]{
    colback=red!5!white,
    colframe=red!75!black,
    coltitle=white,
    colbacktitle=red!75!black, % Background color of title
    title={\strut#1 \hardwarechallenge}, % Strut for consistent height, Wrench symbol
    enhanced,
    attach boxed title to top left={yshift=-2mm, xshift=2mm},
    boxed title style={colframe=red!75!black, colback=red!75!black}, % Style for title background
    left=2mm,
    right=2mm,
    fonttitle=\bfseries,
    halign title=flush center,
}

\title{\sol{}: A Compiler for Neutral Atom Quantum Computers under Hardware Constraints}

\author{\IEEEauthorblockN{Jason Ludmir}
\IEEEauthorblockA{\textit{Rice University}\\ Houston, TX, USA}
\and
\IEEEauthorblockN{Tirthak Patel}
\IEEEauthorblockA{\textit{Rice University}\\ Houston, TX, USA}
}

\begin{document}
\maketitle

\begin{abstract}

Among different quantum computing technologies, neutral atom quantum computers have several advantageous features, such as multi-qubit gates, application-specific topologies, movable qubits, homogenous qubits, and long-range interactions. However, existing compilation techniques for neutral atoms fall short of leveraging these advantages in a practical and scalable manner. This paper introduces \sol{}, a zero-SWAP, scalable, and parallelizable compilation and atom movement scheduling method tailored for neutral atom systems, which reduces high-error operations by 25\% and increases the success rate by 28\% on average compared to the state-of-the-art technique.

\end{abstract}

\begin{IEEEkeywords}
Quantum Compiling, Neutral/Rydberg Atoms
\end{IEEEkeywords}

\section{Introduction}
\label{sec:intro}

Quantum computing holds the promise to speed up computational tasks in diverse sectors, including scientific computing, high-performance computing, and machine learning~\cite{preskill2018quantum,preskill2021quantum}. In the past, superconducting qubits have been at the forefront of quantum technology due to early breakthroughs that leverage our knowledge in classical silicon-based technologies~\cite{arute2019quantum}. Nevertheless, in recent years, neutral atom quantum computers have emerged as a promising alternative~\cite{Sibalic2016-pv,Graham2019-ji,schlosser2023scalable}.

Neutral atom systems have capabilities such as executing multi-qubit gates directly, allowing for application-specific qubit topologies, and the ability to move atoms to reconfigure qubit positions dynamically~\cite{pause2023supercharged,wurtz2023aquila,nogrette2014single,Henriet2020-kl,Barredo2018-ie,bluvstein2022quantum}. Additionally, the homogeneous nature of qubits and their ability for long-range interactions present a pathway to more scalable quantum computing solutions~\cite{Wintersperger2023,Levine2019-bp,PhysRevLett.104.010502,isenhower2010demonstration}.

However, as with superconducting qubits, neutral atoms also face challenges related to the high hardware noise and error rates that increase with the number of operations and execution times of quantum algorithms~\cite{Wintersperger2023,Henriet2020-kl,wurtz2023aquila}. Therefore, it is imperative to leverage compiler techniques to reduce the number of operations and execution times of quantum algorithms before they can run on quantum computers so that the fidelity of their output can be maximized. To achieve this, in this work, we propose a compiler technique, \sol{}\footnote{\sol{} is published in the Proceedings of the ACM/IEEE International Conference for High Performance Computing, Networking, Storage, and Analysis (SC), 2024.}, that effectively leverages the properties of neutral atom systems in novel ways. Before we introduce the contributions of \sol{}, we first provide a brief, relevant background on quantum computation and neutral atom quantum computing technology.

\begin{figure}
    \centering
    \includegraphics[scale=0.55]{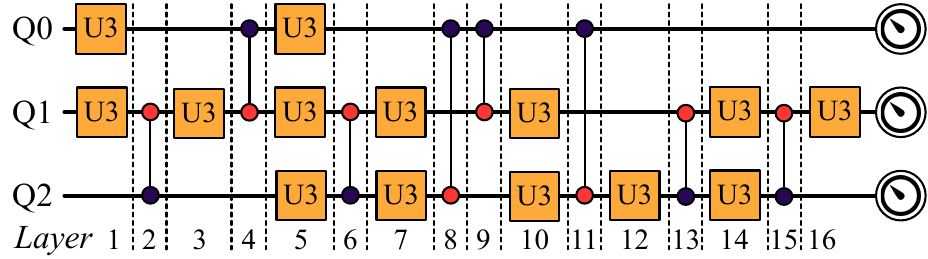}
    \vspace{1mm}
    \hrule
    \vspace{1mm}
    \caption{The depicted Fredkin circuit has three qubits, represented by horizontal lines, on which the one-qubit U3 gates and two-qubit CZ gates are applied (represented by the vertical lines connecting two qubits). The qubits are measured at the end to get the output probability distribution. The circuit has 16 layers. Gates within a layer are parallelly executable.}
    \label{fig:fredkin}
    \vspace{-6mm}
\end{figure}

\subsection{Brief Background of Quantum Technology} %Moved background to its own section, following guidance of reviewer 3

\noindent\textbf{Qubits, Quantum Gates, and Quantum Circuits.} The atomic unit of a quantum computer is the \textbf{qubit}. A qubit can exist in any superposition of the 0 and 1 states, represented as $\ket{\psi} = \alpha \ket{0} + \beta \ket{1}$. Here, $\alpha$ and $\beta$ are complex numbers that represent the magnitudes of the $\ket{0}$ and $\ket{1}$ states. Upon measurement, the probability of measuring a qubit in the $\ket{0}$ state is $\norm{\alpha}^2$, and $\ket{1}$ state is $\norm{\beta}^2$. An entangled state of $n$ qubits is $\ket{\Psi} = \sum_{k=0}^{k=2^n-1}\alpha_k\ket{k}$, such that $\sum_{k=0}^{k=2^n-1}\norm{\alpha_k}^2 = 1$; the probability of observing the $k^{th}$ state is $\norm{\alpha_k}^2$.

\textbf{Quantum operations or gates}, represented as unitary matrices, are used to control qubit states. The one-qubit \textbf{U3} gate is used for superposition, and the two-qubit \textbf{CZ} gate is used for entanglement. Their matrices are as follows.
\vspace{-1mm}
\begin{gather}
	\text{U3} = \begin{bmatrix}
	\text{cos($\frac{\theta}{2}$)} & -e^{i\lambda}\text{sin($\frac{\theta}{2}$)} \\
	e^{i\phi}\text{sin($\frac{\theta}{2}$)} & e^{i(\phi+\lambda)}\text{cos($\frac{\theta}{2}$)}
	\end{bmatrix} \ \
  	\text{CZ} = \begin{bmatrix}
 	1 & 0 & 0 & 0 \\
  	0 & 1 & 0 & 0 \\
  	0 & 0 & 1 & 0 \\
  	0 & 0 & 0 & -1
  	\end{bmatrix}
	\notag
\end{gather}
These two gates form a universal basis, allowing any quantum algorithm to be represented using combinations of these two gates. The \textbf{SWAP} gate, which consists of three CZ gates, is used to exchange the states of two qubits. It is used when two qubits are too far to interact directly. Therefore, their states get swapped with interim qubits to bring them closer.

A \textbf{quantum algorithm or circuit} is a sequence of quantum gates that process information before the qubits are measured. See Fig. \ref{fig:fredkin} for an example of a small quantum circuit.

\vspace{2mm}

\begin{figure}
    \centering
    \includegraphics[scale=0.51]{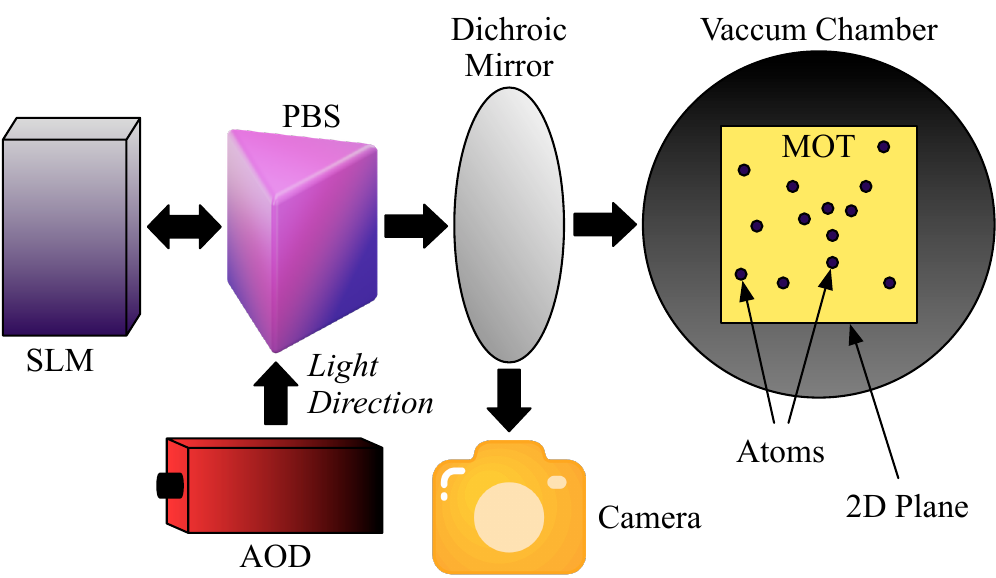}
    \vspace{1mm}
    \hrule
    \vspace{1mm}
    \caption{Device layout of operating a neutral atom quantum computer. The atoms are suspended in a vacuum chamber and controlled using SLM and AOD devices.}
    \label{fig:devices}
    \vspace{-8mm}
\end{figure}

\noindent\textbf{Neutral Atom Quantum Computing.} This technology uses atoms of elements such as Rubidium or Cesium as qubits~\cite{Sibalic2016-pv,Graham2019-ji} -- thus, we use ``qubits'' and ``atoms'' interchangeably. 

Atoms, once cooled in a magneto-optical trap (MOT), are trapped by an acousto-optic deflector (\textbf{AOD}) and a spatial light modulator (\textbf{SLM}) via the polarizing beam splitter (PBS) (Fig. \ref{fig:devices})~\cite{Levine2018-ma,Levine2019-bp,Henriet2020-kl,saffman2020symmetric,barredo2016atom,endres2016atom}. The SLM creates an array of static optical tweezer traps~\cite{pause2023supercharged,schlosser2011scalable,gyger2024continuous,dumke2002micro} in a specified configuration, while the AOD provides a mobile array of optical traps, a feature central to the design principles of \sol{}~\cite{schlosser2012fast,lengwenus2010coherent}. After loading, the SLM displays a grid of atoms, each at a user-defined location. Operations on the qubit-atoms are conducted using specific lasers, and post-computation, a fluorescence-sensing camera reads out the atom array~\cite{Adams2019-sc,schlosser2023scalable}.

The grid of atoms trapped in the SLM and AOD are together called the \textbf{qubit topology or layout}. The topology, therefore, defines how the qubits are positioned in the hardware.

\vspace{2mm}

\begin{figure}
    \centering
    \subfloat[Interaction \& Blockade Radius]{\includegraphics[scale=0.565]{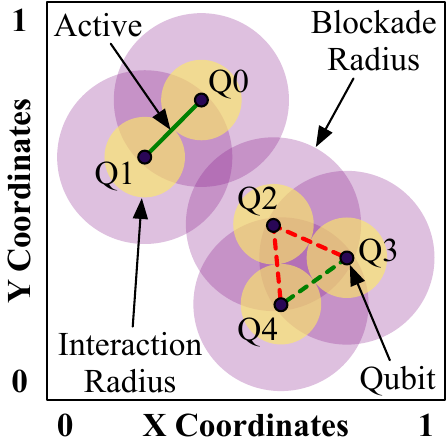}}
    \hfill
    \subfloat[AOD vs. SLM]{\includegraphics[scale=0.565]{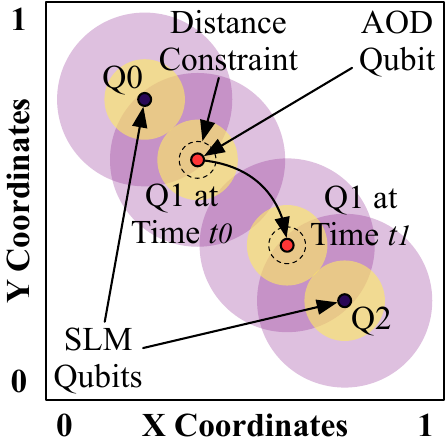}}
    \vspace{1mm}
    \hrule
    \vspace{1mm}
    \caption{(a) When a two-qubit gate is running on Q0-Q1, a two-qubit gate can simultaneously run on Q3-Q4, but not on Q2-Q3 or Q2-Q4, as Q2 is blocked when Q0-Q1 are interacting. The radius of the circles corresponding to the interaction radius represents half of the actual interaction radius. We draw the half-radius for ease of interpretation. Thus, if two qubits' interaction radii circles touch or overlap, they can interact with each other. Similarly, the radius of the circles corresponding to the blockade radius represents half of the actual blockade radius. (b) Qubits on the SLM are stationary throughout circuit execution, while qubits on the AOD are mobile. The radius of the circles corresponding to the distance constraint represents half of the minimum separation distance constraint.}
    \label{fig:ryd_aod_slm}
    \vspace{-6mm}
\end{figure}

\noindent\textbf{Gate Implementations.} Computation is performed via different sets of lasers interacting with the atoms in the AOD/SLM arrays. \textbf{Raman transitions}, which involve pulses from lasers to transition an atom between the $\ket{0}$ and $\ket{1}$ hyperfine states, are used to perform the U3 gate with arbitrary rotation~\cite{Henriet2020-kl}.
One of the main benefits of neutral atom systems is the ability to connect distant atoms using their \textbf{Rydberg states}. When an atom is highly excited, its outermost electron occupies a Rydberg state, meaning it has an enlarged orbit that allows interactions with nearby atoms. Local Rydberg lasers can be used to excite individual atoms~\cite{PhysRevA.90.023415}. These long-range atom interactions are used to implement CZ gates on atoms within the interaction radius of each other~\cite{Levine2019-bp,PhysRevLett.104.010502,isenhower2010demonstration}.
While this allows for interactions between atoms, it also results in the \textbf{Rydberg blockade effect}: If one atom is in a Rydberg state, neighboring atoms within a certain radius cannot be similarly excited if they are not involved in a gate with that atom. This blockade effect restricts parallelism since any atoms in the blockade radius cannot themselves perform multi-qubit gates (though they can still perform single-atom Raman transitions). This blockade radius is typically $2.5\times$ the interaction radius~\cite{tan2024compiling}. See Fig.\ref{fig:ryd_aod_slm}(a) for an example of how Rydberg atoms interact with blockades. 

Despite these constraints, different gates can be executed in parallel on neutral atom systems. Both one-qubit U3 gates and two-qubit CZ gates can be executed in parallel on different qubits in the system (but not on the same qubit in a single layer). Notably, single qubit Raman transitions (used for U3 gates) have been demonstrated to occur simultaneously with the Rydberg excitations used to execute CZ gates~\cite{regional_addressing,radnaev2024universal}.

\vspace{2mm}

\noindent\textbf{Atom Movements and Measurements.} Recall that qubits can be trapped by both the SLM (static) and AOD (mobile). The AOD qubits can be repositioned dynamically as the program progresses, as shown in Fig.~\ref{fig:ryd_aod_slm}(b). This allows for qubits that are not within the Rydberg interaction radius to be moved into range of each other without the need to insert SWAP operations to bring them closer. However, there are several restrictions to moving atoms via the AOD.

First, each atom has a minimum separation \textbf{distance constraint}, whereby it must never be closer to any other atom by some hardware-dependent distance~\cite{wurtz2023aquila,huang2024dasatom,beugnon2007two}. Second, the AOD grid is comprised of rows and columns of traps. AOD rows/columns cannot cross over each other, meaning the relative order of the rows and columns must be maintained for the duration of the program~\cite{bluvstein2022quantum}. This is due to the fact that the AOD traps might interfere with one another if they are moved across each other. Lastly, qubits on an AOD row/column have to move in tandem~\cite{tan2022qubit,bluvstein2022quantum}. This means if one row has two atoms on it, and one atom is moved up by $x$ units, the other atom is moved up by $x$ as well.

As quantum computing is probabilistic, the same circuit has to be run and measured multiple times to construct the output probability distribution. Each such run is referred to as a \textbf{logical shot}. When multiple logical shots are run together, in parallel, in one hardware shot, we call this a \textbf{physical shot}.

\vspace{2mm}

\noindent\textbf{Hardware Noise and Error Effects.} Noise in neutral atom quantum systems arises from a variety of sources, which can lead to erroneous output. Atoms in superposition states naturally decohere over time, i.e., lose their state. While atoms stay coherent in a hyperfine state for around 1-3 seconds, they stay coherent in a Rydberg state for only around 5-10 $\mu{}s$~\cite{Wintersperger2023,bluvstein2022quantum}. Another source of noise is the readout error, whereby the fluorescent camera incorrectly reads the measured state of a qubit. Readout error on current systems is around 5\%~\cite{Wintersperger2023}.

Accidental atom loss, where atoms escape from optical traps, occurs at a rate of about 0.7\%~\cite{bluvstein2022quantum}. This can occur due to collisions of free atoms in the MOT with trapped atoms in the AOD and SLM~\cite{Henriet2020-kl,Baker2021-wv}. Another source of error is operational error, which refers to errors introduced during gate operations on qubits. Different quantum gates inherently possess varying error rates. For example, the operational error associated with single-qubit U3 gates is relatively low, with studies indicating an error rate of approximately 0.01\%~\cite{PhysRevA.105.032618}.

In contrast, CZ gates exhibit a significantly higher operational error rate, around 0.5\%~\cite{highfidbluv}. This highlights the challenge with SWAP operations, which are composed of three CZ gates, thus having an error of 1.43\%. As a result, avoiding SWAP operations is a major motivator of \sol{}.

\begin{table}[t]
    \centering
    \caption{Comparison of the functionalities of different neutral atom compilation works. Only \sol{} achieves all functionalities in a practical and scalable manner.}
    \vspace{-1mm}
    \scalebox{0.82}{
    \begin{tabular}{cccccc}
         \textbf{Technique} & \textbf{Practical} & \textbf{Custom} & \textbf{Atom} & \textbf{Zero} & \textbf{Parallel Shot} \\
          & \textbf{\& Scalable} & \textbf{Layout} & \textbf{Movement} & \textbf{SWAPs} & \textbf{Movements} \\
         \\[-2.4mm]
         \hline
         \hline
         \\[-2.4mm]
         \eldi{}~\cite{Baker2021-wv,litteken2022reducing} & \cmark{} & \xmark{} & \xmark{} & \xmark{} & \xmark{} \\
         \\[-2.4mm]
         \hline
         \\[-2.4mm]
         \geyser{}~\cite{patel2022geyser} & \cmark{} & \xmark{} & \xmark{} & \xmark{} & \xmark{} \\
         \\[-2.4mm]
         \hline
         \\[-2.4mm]
         \graphine{}~\cite{patel2023graphine} & \cmark{} & \cmark{} & \xmark{} & \xmark{} & \xmark{} \\
         \\[-2.4mm]
         \hline
         \\[-2.4mm]
         \olsq{}~\cite{tan2022qubit,tan2024compiling} & \xmark{} & \cmark{} & \cmark{} & \cmark{} & \xmark{} \\
         \\[-2.4mm]
         \hline
         \hline
         \\[-2mm]
         \sol{} & \cmark{} & \cmark{} & \cmark{} & \cmark{} & \cmark{} \\
         \\[-2.6mm]
         \hline
         \hline
    \end{tabular}}
    \vspace{-4mm}
    \label{tab:results}
\end{table}

\subsection{Related Work and Limitations}

Table~\ref{tab:results} shows the functionalities of several state-of-the-art compilers for neutral atom quantum computers. Baker et al.~\cite{Baker2021-wv} propose a mapping and routing scheduler for gates to run on neutral atoms arranged in a square grid -- this work is referred to as \eldi{} in this paper. An extension of this work~\cite{litteken2022reducing} explores the trade-off between remapping/recompiling strategies between shots and parallelization. Further, \geyser{}~\cite{patel2022geyser} composes smaller gates into multi-qubit gates for efficient execution on neutral atoms arranged in a triangular grid. \geyser{} is orthogonal to \sol{} as it is a gate composer that can be used in conjunction with any compilation technique. Both of the above works do not have custom layouts or atom movements.

On the other hand, \graphine{}~\cite{patel2023graphine} supports custom layouts but not atom movements. Note, similar to \graphine{}, a work proposed by Nguyen et al.~\cite{nguyen2023quantum} designs custom neutral atom layouts, but only for combinatorial optimization applications. Since \graphine{} is the general solution for designing layouts, we use it as a comparison point for \sol{}. In contrast to \graphine{}, \olsq{}~\cite{tan2022qubit,tan2024compiling} supports atom movements, but it assumes global addressing and is not scalable/generalizable. This is because \olsq{} creates an optimal movement plan for atoms using a satisfiability modulo theories (SMT) model, which must take into account every one of the constraints on AOD and atom movement, Rydberg interaction and blockade radii, and the structure of the input circuit. This leads to prohibitively large compilation times -- we were not able to compile even small 9-16 qubit circuits within 24 hours.

Owing to the above reasons, we use \eldi{} and \graphine{} as state-of-the-art comparison points for \sol{}.

Note that the above works do not make a practical and scalable use of all of the neutral atom properties. Specifically, previous techniques, including the ones that support atom movement, still use many high-error SWAP or time-consuming atom release/retrap operations. Eliminating these operations is a non-trivial problem as it requires the circuit execution schedule to be designed such that they rarely become necessary. Previous work also does not combine atom movement with parallel shot execution; unlike superconducting-qubit architectures, neutral atom architectures have homogeneous qubits -- all qubits have the same quality, and thus there is no requirement to only run a circuit on the best low-error qubits, which is typically done for superconducting qubits~\cite{tannu2019ensemble,wille2019mapping,li2022paulihedral}. The ability to utilize all the atoms can be used to carefully parallelize multiple logical shots while allowing atom movement, thus reducing the total time of executing all shots.
\subsection{Contributions of \sol{}}%moved from intro, also in line with Rev. 3's recs

\begin{figure*}[t]
    \centering
    \includegraphics[scale=0.53]{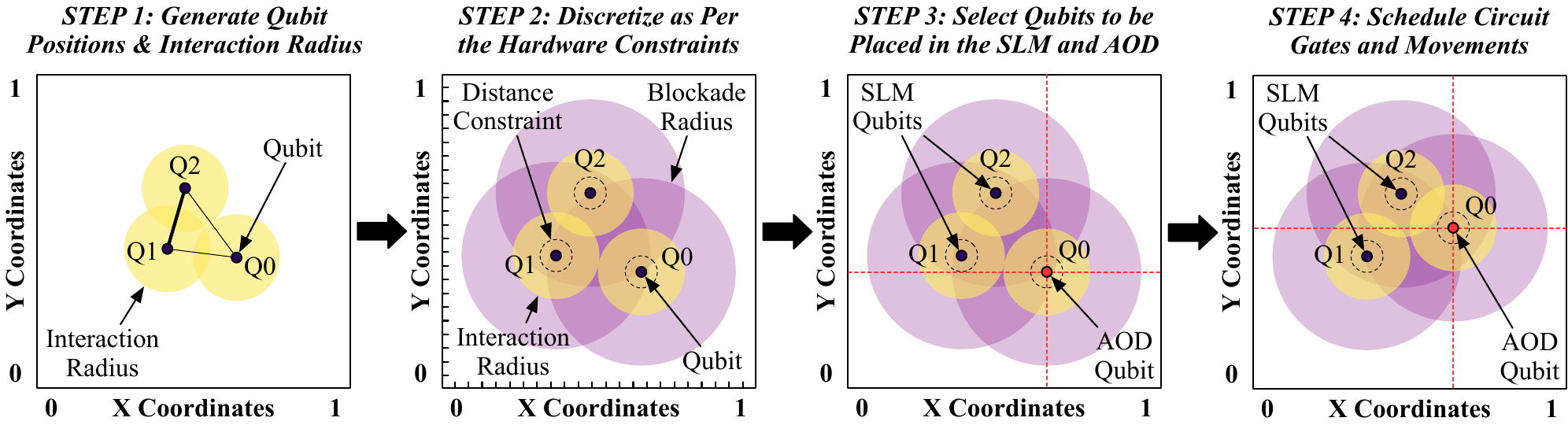}
    \vspace{1mm}
    \hrule
    \vspace{1mm}
    \caption{A high-level overview of the four steps that \sol{} takes to compile a circuit for neutral atom quantum computers. The steps here correspond to the three-qubit Fredkin circuit shown in Fig.~\ref{fig:fredkin}. The atom configuration in Step 3 corresponds to layers 1-7 and 9 in the Fredkin circuit, and the one in Step 4 corresponds to layers 8 and 10-16 in the Fredkin circuit.}
    \label{fig:steps}
    \vspace{-6mm}
\end{figure*}

To overcome the above limitations, we design \sol{}, the first compiler to leverage qubit homogeneity and atom movement to support zero-SWAP and highly parallelizable circuit executions. \sol{} designs atom movements in such a way that it becomes possible to move atoms simultaneously for all parallel circuit shots. It does so by running a four-step procedure that (1) carefully initializes qubit positions, (2) discretizes them under hardware constraints, (3) selects only a small subset of qubits to be trapped by the AOD, and (4) carefully schedules gates and atom movements using scalable and parallelizable heuristics.

\vspace{2mm}

\noindent\textbf{The contributions of this work are as follows:}
\begin{itemize}
    \item This work introduces \sol{}, the first heuristical compilation method to have zero SWAP gates and parallelize circuit executions while facilitating atom movements and satisfying nontrivial hardware constraints.
    \item \sol{} uses a combination of optimal topology creation, heuristic-based gate scheduling, and recursive atom movement to eliminate the need for SWAP gates.
    \item \sol{} enables the replication and execution of multiple circuits simultaneously on a larger atom grid, effectively addressing scalability issues of atom movement by maximizing the utilization of the system's hardware.
    \item \sol{}'s open-source and parallel implementation shows that by eliminating SWAPs, \sol{} reduces the number of CZ gates in quantum circuits by 39\% over \graphine{}~\cite{patel2023graphine} and 25\% over \eldi{}~\cite{Baker2021-wv,litteken2022reducing} on average. 
    \item \sol{} improves the probability of success by 46\% over \graphine{} and 28\% over \eldi{}, while achieving similar runtimes on average. \sol{}'s implementation and data are available at: \texttt{\url{https://github.com/positivetechnologylab/Parallax}}.
\end{itemize}
\section{Design and Implementation}
\label{sec:design}

\begin{figure}
    \centering
    \vspace{-2mm}
    \subfloat[Distance Constraint Rationale]{\includegraphics[scale=0.57]{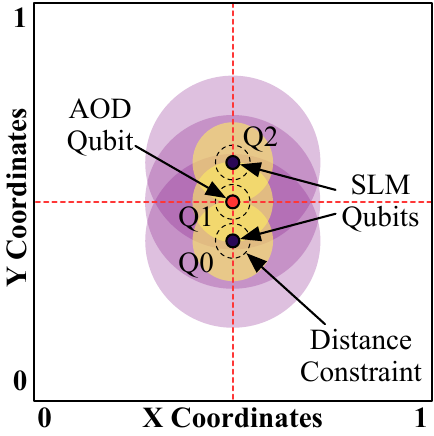}}
    \hfill
    \subfloat[One AOD per Row/Column]{\includegraphics[scale=0.57]{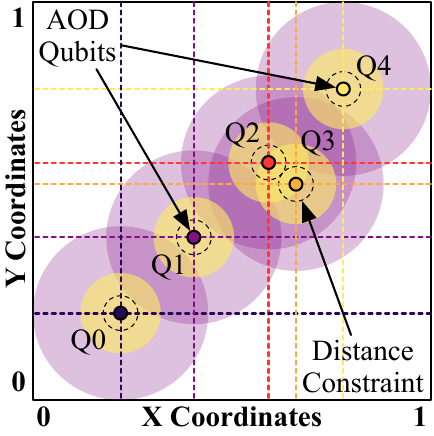}}
    \vspace{1mm}
    \hrule
    \vspace{1mm}
    \caption{(a) \sol{} ensures there is enough gap among SLM qubits for AOD qubits to traverse through. (b) \sol{} ensures one qubit per row/column in the AOD.}
    \label{fig:aod}
    \vspace{-7mm}
\end{figure}

In this section, we describe the design of \sol{} to ensure zero SWAP operations and high parallelism of logical shots. These steps are visually depicted in Fig.~\ref{fig:steps}. First, \sol{} generates a custom qubit layout for a given circuit using a method similar to \graphine{}~\cite{patel2023graphine}. It then discretizes the qubit location given the hardware constraints and optimizes the placement of qubits into static SLM and mobile AOD traps. Lastly, it uses several heuristics to schedule gates and atom movements during the parallel execution of the circuit.

Next, we discuss all of these steps in detail.
 
\subsection{Initialization of Qubit Placement}

Before executing the circuit, \sol{} must first load atoms into the SLM, which creates static atom traps. A starting atom topology where atoms are placed close to other atoms they interact with would be desirable. This is because atom movement by the AOD is highly constrained: if atoms that interact frequently are placed far apart, there will be a significant amount of unnecessary movement to bring them close together, leading to higher run time and error. Thus, \sol{} needs to produce an initial atom topology where frequently interacting atoms are placed close by.

\sol{} uses \graphine{} to construct this initial topology. \graphine{} is a method for generating approximately optimal static topologies for neutral atom hardware~\cite{patel2023graphine}. Employing \graphine{}, we first convert the input circuit into a graph by representing the qubits as nodes and the number of gates between any two qubits as weighted edges. Next, we use \graphine{}'s dual annealing~\cite{sahin1998dual} to place qubits on a 2D plane. Dual annealing is a global optimization algorithm that employs a broad search of the entire solution landscape, gradually focusing on promising areas. As it identifies locations of where global optima might reside, it slowly switches to a more precise, local search to refine the most potential solutions.
The annealer is optimized to place pairs of qubits with high-weight edges closer together since this implies many shared gates. Then, we use \graphine{} to select a Rydberg interaction radius large enough to ensure that all of the qubits are reachable from all other qubits, i.e., the qubits form a connected graph, and no qubit is isolated from the rest.

Note that the generated qubit locations do not respect many hardware constraints. Recall that \sol{} must ensure that atoms are spaced out enough to respect the minimum separation distance constraint. Additionally, \sol{} must place atoms far enough apart to prevent them from forming barriers that could impede other atoms' movement.

\vspace{2mm}

\noindent\textbf{\textcolor{green}{\solution} Solution:} \sol{} discretizes the atom array such that a unit of discretization represents twice the minimum separation distance, plus a small amount of padding to allow for atoms to navigate around each other (see Fig. \ref{fig:steps} Step 2). There are two rationales for this. (1) This guarantees the initial atom topology will not violate the minimum separation distance constraint of the atoms. (2) When moving atoms, there will be guaranteed space to pass between stationary atoms in the SLM, as shown in Fig.~\ref{fig:aod}(a). By ensuring there is ample space around each static SLM atom, \sol{} not only adheres to the distance constraint but also facilitates seamless navigation for mobile atoms in the system, which we discuss next. 

\subsection{Planning for Optimal AOD Movement}

\begin{figure}
    \centering
    \includegraphics[scale=0.52]{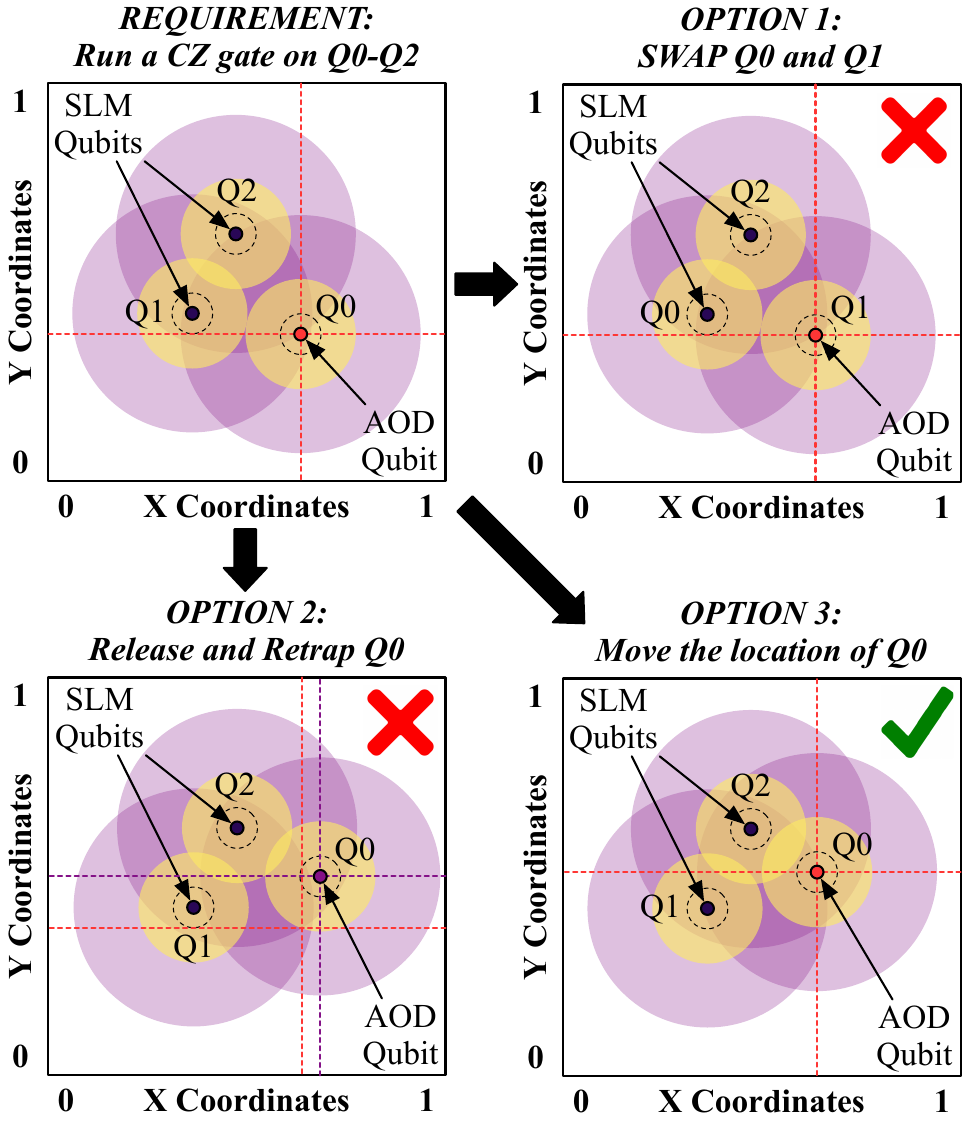}
    \vspace{1mm}
    \hrule
    \vspace{1mm}
    \caption{\sol{}'s design ensures that it can move AOD qubits as frequently as possible, which helps it reduce the error and runtime (as verified by \sol{}'s evaluation in Sec.~\ref{sec:eval}). Example corresponds to the circuit in Fig.~\ref{fig:fredkin}.}
    \label{fig:options}
    \vspace{-6mm}
\end{figure}

The ability to move atoms physically using the AOD is a key advantage of neutral atom systems. Atoms confined to static traps in the SLM can be transferred to the AOD, where they become movable. Consequently, although only the atoms in the AOD can be moved, the interchange of atoms between the SLM and the AOD ensures that all atoms can be mobile. Thus, when a neutral atom system attempts to execute a two-qubit gate where the atoms are out of the Rydberg interaction radius, it has three options, as shown in Fig. \ref{fig:options}. \textbf{Option 1.} If neither atom is in the AOD, the system can perform SWAP operations to move the atoms into the interaction radius of each other. However, this has a major downside. SWAP operations require three CZ gates to execute, and a single SWAP gate has an error rate of roughly $1.43\%$~\cite{highfidbluv}. Even a small number of SWAPs can thus introduce a prohibitive amount of error into a circuit. \textbf{Option 2.} If neither atom is in the AOD, the system can trap one of the atoms from the SLM into the AOD and then move it into the interaction radius. While this method has a relatively low error rate, it has a high time cost, with an estimated 100$\mu$s to complete trap switches~\cite{tan2024compiling}. \textbf{Option 3.} If at least one of the atoms is already in the AOD, it can move the AOD atom into the interaction radius of the other atom. It has been shown that atom transport is both low error ($<$0.1\% atom loss from movement) and fast (average speed of 55$\mu$m/$\mu$s)~\cite{bluvstein2022quantum}.

Given the above options, movement with atoms already in the AOD is the preferable method to get atoms into position to execute CZ gates. Therefore, we design \sol{} to use Option 3 as often as possible. However, moving an atom with the AOD requires one of the atoms involved in the gate to be already in the AOD. Moreover, atom movement has a number of constraints that complicate compilation. (1) AOD rows and columns cannot cross over each other, as optical trap frequencies get garbled up. (2) All of the atoms on an AOD row/column move in tandem. (3) Atoms are subjected to the aforementioned minimum separation distance constraint.

Further, \sol{} must also optimize to achieve the best placement of atoms for upcoming operations and the ideal movement patterns for rows and columns to minimize overlap.

\vspace{2mm}

\noindent\textbf{\textcolor{green}{\solution} \ Solution:} \sol{} considerably reduces the complexity of creating an AOD movement plan by placing only one atom per AOD row/column pair. This achieves the following: (1) Rows and columns cross over each other with significantly lower frequency, especially since \sol{} localizes each atom close to other atoms it interacts with. (2) No other atoms have to move in tandem with atoms in the same row and column. This increases the average number of gates we can execute in parallel since atoms are never moved out of the interaction radius of another atom for the sake of some other atom on the row/column. (3) There is less obstruction from the atoms' minimum separation distance constraint as atom collision is also less frequent. Recall that \sol{} ensures that SLM atoms are easy to avoid since they are static and always have space between them; AOD atoms are relatively few in number and can be moved out of the path of any other AOD atom (see Fig. \ref{fig:aod}(b)). (4) The simplicity of having only one atom for each row/column pair will make it easier to parallelize \sol{}'s different operations.

One might ask why we do not simply place all atoms in either the SLM or AOD. If all atoms are placed in the SLM, SWAPs would be the only option for bringing atoms closer, increasing error and execution time. If all atoms are placed in the AOD, the separation distance and AOD row/column overlap constraints become so prohibitive that a large number of SWAP operations and trap changes would be required, which would again increase error and execution time.

\subsection{AOD Qubit Selection}

Next, we discuss how \sol{} selects qubits to place in the AOD while maintaining one qubit per AOD row/column.

\vspace{2mm}

\noindent\textbf{\textcolor{green}{\solution} Solution:} To select mobile AOD atoms, \sol{} uses a heuristic algorithm that weighs each atom by the following two criteria: (1) the number of times an atom interacts with other atoms outside of its interaction radius, and (2) the degree of serialization that would be caused by the atom's blockade radius. The first consideration is paramount since two interacting atoms being out of each other's interaction radius will necessitate a move of one atom closer to the other. If neither is in the AOD, the AOD will need to trap one of the atoms mid-computation or perform one or more SWAPs. To avoid this, \sol{} includes one of these atoms in the AOD. By default, this is weighted at 0.99 out of 1.

The other criterion used to weigh is how often a qubit interferes with other multiqubit gates in the same layer. The more interference it causes, the more serialization of individual layers it will need to perform if it cannot move the qubit. This metric is of lower importance and is used as a tie-breaker between qubits with the same number of out-of-range interactions. Therefore, this is weighted at 0.01 out of 1. The set of highest-weight qubits are placed in the AOD as close to their initial locations as possible while respecting the rules for AOD movement. This step is shown in Fig.~\ref{fig:steps} Step 3.

However, this creates a new challenge. The atoms all start in the SLM, where atoms can be in the same row or column without issue. However, once the atoms are placed in the AOD, atoms cannot share row and column coordinates to comply with \sol{}'s aforementioned requirement to only place one qubit in each row/column. \sol{} must address this.

\vspace{2mm}

\noindent\textbf{\textcolor{green}{\solution} Solution:} For each AOD row/column, we check if it shares a position with another AOD row/column. If it does, \sol{} will move it a small amount in a chosen direction (e.g., for rows, always move the rows up). \sol{} then recurses on other rows or columns if, after movement, there are shared positions such that at the end of the process, no row or column occupies the same position.

Now that \sol{} has selected locations for the atoms and the devices that the atoms are trapped in (SLM vs. AOD), it is time for \sol{} to schedule gates for execution.

\begin{algorithm}[t]
\caption{\sol{}'s gate scheduling algorithm.}
\label{alg:compile_circ}
\begin{algorithmic}[1]
\State $G \gets \text{List of un-executed gates in circuit}$
\State $Q \gets \text{List of qubits}$
\State $r \gets \text{Rydberg interaction radius}$
\State $c \gets \text{number of gates before repositioning AOD}$
\State \textsc{Compile\_Circuit}($G,Q,r,c$)
\State \ \ \ \textbf{while} Gates still left in $G$ \textbf{do}
\State \ \ \ \ \ \ Initialize empty $curr\_layer$ list
\State \ \ \ \ \ \ \textbf{for} each qubit $q$ in $Q$ \textbf{do}
\State \ \ \ \ \ \ \ \ \ \textbf{if} $q$'s dependencies are satisfied \textbf{then}
\State \ \ \ \ \ \ \ \ \ \ \ \ Get the next gate $g$ for $q$
\State \ \ \ \ \ \ \ \ \ \ \ \ Append $g$ to $curr\_layer$
\State \ \ \ \ \ \ \textbf{for} CZ gates in $curr\_layer$ \textbf{do}
\State \ \ \ \ \ \ \ \ \ \textbf{if} gate $g$ is out of range \textbf{then}
\State \ \ \ \ \ \ \ \ \ \ \ \ \textbf{if} qubit $g_{q1}$ in AOD \& no moves yet \textbf{then}
\State \ \ \ \ \ \ \ \ \ \ \ \ \ \ \ Move $g_{q1}$ to the interaction radius $g_{q2}$
\State \ \ \ \ \ \ \ \ \ \ \ \ \textbf{else if} already moved in $curr\_layer$ \textbf{then}
\State \ \ \ \ \ \ \ \ \ \ \ \ \ \ \ Pop $g$ from $curr\_list$ and add it back to $G$
\State \ \ \ \ \ \ \ \ \ \ \ \ \textbf{else if} neither qubit in $g$ is in the AOD \textbf{then}
\State \ \ \ \ \ \ \ \ \ \ \ \ \ \ \ Trap and move $g_{q1}$
\State \ \ \ \ \ \ Shuffle elements of $curr\_layer$
\State \ \ \ \ \ \ \textbf{for} CZ gates in $curr\_layer$ \textbf{do}
\State \ \ \ \ \ \ \ \ \ Modify $curr\_layer$ s.t. no interference occurs
\State \ \ \ \ \ \ Execute gates in $curr\_layer$
\State \ \ \ \ \ \ Reset positions of moved atoms
\end{algorithmic}
\end{algorithm}
\subsection{Scheduling of Gates and Movements}

To schedule gates, \sol{} must consider the following. (1) Gates must execute in order (i.e., dependencies must be preserved). (2) If two atoms share a gate, they must be moved within the interaction radius of each other if not already. And (3) gates that are in the same layer that interfere with each other's execution must be serialized.

We now give a detailed explanation of Algorithm~\ref{alg:compile_circ}, which schedules gates given the above requirements. When executing a quantum circuit, gates are performed on qubits sequentially. The gates that together compose a quantum circuit are not necessarily commutable; thus, the order of the gates that operate on each qubit must be preserved while ensuring that the compiled circuit maximizes parallelism.

\vspace{2mm}

%Solution
\noindent\textbf{\textcolor{green}{\solution} Solution:} \sol{} builds layers of gates that can operate in parallel. For each layer, it attempts to add one gate per qubit, finding the next gate each qubit can execute based on the dependency graph. One-qubit gates are simple to add since they only depend on gates for the corresponding qubit. The two qubits could have complex dependencies, and thus, one qubit might have to be stalled until the other qubit involved in the gate ``catches up''. See lines 8-11 in Algorithm~\ref{alg:compile_circ}. 

Once the layers have been constructed, \sol{} must schedule any atoms that are involved in a CZ operation together that are out of the interaction radius of each other. 

\vspace{2mm}

\noindent\textbf{\textcolor{green}{\solution} Solution:} This is where \sol{} utilizes AOD movement: if at least one of the two out-of-radius atoms is in the AOD, \sol{} moves the AOD atom within the interaction distance of the other atom. Note that at least one atom in an out-of-radius interaction is highly likely to be in the AOD since \sol{} selected AOD atoms mainly based on the number of out-of-radius operations. See lines 12-19 in Algorithm~\ref{alg:compile_circ}.

Whenever an atom needs to be moved, \sol{} performs a recursive move to get the atom from its starting point to its endpoint. If an atom moves to be within the minimum separation distance of another AOD atom, the system will recursively move that obstructing atom out of the way and keep recursing on AOD atoms that obstruct other moved atoms until sufficient room exists for the first atom to be within the interaction radius of the atom it is interacting with. Similarly, if as the atom moves, its AOD row and column get too close to other AOD rows or columns, \sol{} moves the interfering rows and columns recursively away. Trying to move different atoms in the same layer might lead to infinite recursion since the moving atoms might be blocking each other. Thus, in any given parallel layer of gates, only one move-into-range is allowed to occur. This single-move-per-layer paradigm also enables simpler AOD movement schemes, which will make parallelizing circuits easier (discussed in Sec.~\ref{parallel_log}).

Note that there is never an issue of static SLM atoms completely blocking movement -- they might obstruct movement, in which case the moving atom simply needs to move around the obstruction, but due to how \sol{} discretizes the atom grid, there is always room for AOD atoms to move around. Also note that in the rare case that two atoms are both in the SLM and out-of-radius for a CZ gate, \sol{} has no choice but to trap and move one of the atoms. Empirically, we observe that this takes place for 1.3\% of CZ gates across all circuits we have evaluated (Sec.~\ref{sec:eval}).

Now that we have a movement schedule for the atoms in this layer, \sol{} must measure the distances between all atoms involved in CZ gates in that layer to check for Rydberg blockade effects. However, if \sol{} always chooses the same order of atoms to compare distances (e.g., beginning with the highest qubit index), it could lead to situations where gates involving certain atoms get pushed back in execution repeatedly, leading to sub-optimal scheduling. 

The main cause of this sub-optimality comes from the Rydberg blockade interference check: when gates in the same layer interfere with each other, one of the interfering gates must be returned to the unexecuted gate list $G$ to be executed in another layer. If one particular qubit's gate keeps getting pushed back due to the blockade effect, the critical path of the compiled circuit might increase, leading to a longer circuit runtime. To alleviate this, the algorithm shuffles the gate elements of the current layer before continuing. By shuffling the gate order in the layer, \sol{} reduces the chance that one qubit gets arbitrary execution preference over the others. See line 20 in Algorithm~\ref{alg:compile_circ}. Once the gates of the current layer have been shuffled, we must check for Rydberg Blockading in the layer. Atoms that are in Rydberg Blockade radius of each other cannot execute multi-qubit gates in parallel.

\begin{figure}
    \centering
    \vspace{-2mm}
    \includegraphics[scale=0.525]{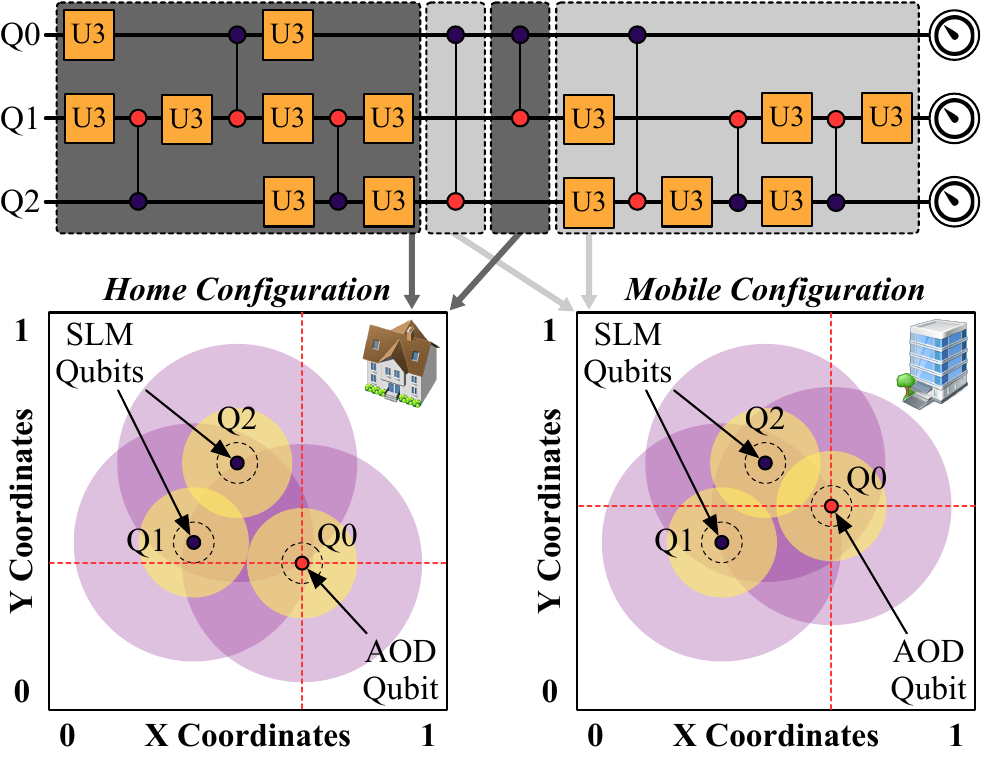}
    \vspace{1mm}
    \hrule
    \vspace{1mm}
    \caption{As shown for the example Fredkin circuit in Fig.~\ref{fig:fredkin}, \sol{}'s AOD qubits travel between home and mobile configurations based on the gates being executed.}
    \label{fig:homing}
    \vspace{-6mm}
\end{figure}

\vspace{2mm}
%Solution
\noindent\textbf{\textcolor{green}{\solution} Solution:} If there are a pair of gates that are interfering with each other, \sol{} ejects one of them out of the current layer and back to the unexecuted gate list, to be executed in a subsequent layer. See lines 21-22 in Algorithm~\ref{alg:compile_circ}.

After checking dependencies, moving atoms, and checking for Rydberg blockade effects, \sol{} has constructed a parallelized layer whose gates can be executed together. However, there is an issue here: after execution, the moved AOD atoms are no longer in their initial ``home'' locations. Before moving, AOD atoms were placed in optimal locations such that they were near other atoms they shared gates. Thus, \sol{} returns the AOD atoms from their ``mobile'' locations back to their positions before the current layer was executed. This is accomplished by reversing the directions of moves. See line 24 in Algorithm~\ref{alg:compile_circ} and Fig. \ref{fig:homing}.

After resetting the locations of the AOD atoms, we have completed the scheduling and execution cycle of a single layer. This process is repeated until all gates are executed. Having outlined the process of scheduling and executing a single circuit, we now turn our attention to an essential aspect of \sol{} not yet discussed: circuit parallelism. This concept is key to optimizing \sol{}'s efficiency and scalability.

\begin{figure}
    \centering
    \includegraphics[scale=0.8]{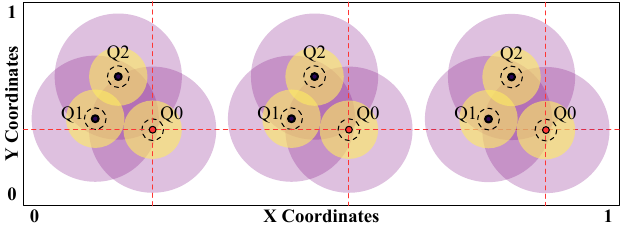}
    \vspace{1mm}
    \hrule
    \vspace{1mm}
    \caption{\sol{} ensures that multiple logical shots can run simultaneously in the same physical shot to maximize qubit utilization. Example corresponds to the circuit in Fig.~\ref{fig:fredkin}.}
    \label{fig:parallel}
    \vspace{-6mm}
\end{figure}

\subsection{Parallelization of Logical Shots}
\label{parallel_log}

We have now shown that using an AOD where each row/column traps one atom, \sol{} compiles quantum circuits in a fast and straightforward manner. Next, we look at how we can parallelize circuits for further improvement.

\vspace{2mm}

\noindent\textbf{\textcolor{green}{\solution} Solution:} In order to maximize usage of the given hardware, \sol{} creates copies of the circuit along the entirety of the system's atom array. Each of these circuits has its own atoms, but they share AOD rows and columns. The shared rows and columns can operate in tandem across all circuits since if the system begins execution of all parallel circuits simultaneously, then all circuits share the same AOD movement scheme. Thus, \sol{} can handle an increased number of atoms on each AOD row and column, maximize the utilization of the system's hardware, and raise the number of shots run per unit of time. See Fig.~\ref{fig:parallel} for an example. In the figure, the row now has three atoms due to \sol{}'s parallelism (by a factor of three in this example). The ability for \sol{} to scale any-sized circuit to any hardware that has enough atoms to run at least one copy of the circuit makes it highly versatile to different circuits and hardware sizes. %Consider two instances. If \sol{} compiles a small 9-qubit circuit for a large 1,225-atom machine, it can simultaneously run 121 copies (11x11) of the circuit, allowing it to maximally utilize the hardware. On the other hand, \sol{} will still optimally compile circuits that might barely fit in one copy of the hardware's atom array. For example, a 255-qubit circuit in a 256-qubit array will still be compiled as effectively as possible. This adaptability in \sol{}'s design not only ensures that hardware resources are never wasted but also allows for a wide range of circuit complexities to be proficiently compiled. 

Next, we discuss \sol{}'s evaluation methodology.

\section{Experimental Methodology}
\label{sec:method}

\noindent\textbf{Experimental Setup.} \sol{}'s evaluation is conducted via a simulator that emulates the hardware characteristics of real neutral atom testbed systems. Note that while a physical 256-qubit neutral atom system does exist (QuEra Aquila~\cite{wurtz2023aquila}), it only supports analog computation and does not yet support a programmable AOD. Thus, to evaluate \sol{}, we developed a simulator using real hardware parameters (see Table~\ref{tab:params}). T1 and T2 times represent the decoherence speed for hyperfine atomic states; we get the decoherence error rates by inputting these times and each benchmark's runtime into exponential decay functions. Note: if atoms are lost during a simulated shot because of atom collision or trap escape (modeled as being part of $T1$ error in our simulation), these atoms are replenished between physical shots as needed. Thus, the impact of atom loss is only on the error rates of the circuits. We use two different hardware simulations for our experimentation. We use QuEra's 256-qubit Aquila system~\cite{wurtz2023aquila}, which has a 16$\times$16 atom grid, for our main results, and Atom's 1,225-qubit system~\cite{norcia2024iterative,russell2023atom}, which has a 35$\times$35 atom grid, to evaluate scaling and parallelization. We use 20 AOD rows and columns as the default configuration. We also ablate it in the next section and show that 20 provides the best results.\begin{table}[t]
    \centering
    \caption{Hardware parameters used for our evaluation.}
    \vspace{-1mm}
    \scalebox{0.92}{
    \begin{tabular}{cc|cc}
         \textbf{Parameter} & \textbf{Value} & \textbf{Parameter} & \textbf{Value} \\
         \hline
         \hline
         Number of Qubits & 256 \& 1,225 & Atom Loss Rate & 0.7\%~\cite{bluvstein2022quantum} \\
         Time to Switch Traps & 100$\mu$s~\cite{tan2024compiling} & U3 Gate Error & 0.0127\%~\cite{PhysRevA.105.032618} \\
         AOD Movement Speed & 55$\mu$m$/\mu$s~\cite{bluvstein2022quantum} & U3 Gate Time & 2$\mu$s~\cite{Wintersperger2023} \\
         T1 Coherence Time & 4.0s~\cite{bluvstein2022quantum} & CZ Gate Error & 0.48\%~\cite{highfidbluv} \\
         T2 Coherence Time & 1.49s~\cite{bluvstein2022quantum} & CZ Gate Time & 0.8$\mu$s~\cite{bluvstein2022quantum} \\
         SWAP Gate Error & 1.43\%~\cite{highfidbluv} &  Readout Error & 5\%~\cite{Wintersperger2023} \\
         \hline
         \hline
    \end{tabular}}
    \vspace{-2mm}
    \label{tab:params}
\end{table}

\sol{} uses a general model to simulate neutral atom computer operations, with objects representing the AOD, SLM, and individual atoms. Given an input circuit in QASM format, \sol{} first executes Graphine to optimize qubit placement using dual annealing, returning coordinates in a [0,1] range for each qubit.  Users can alternatively load pre-obtained Graphine results via a command line argument to reduce compile time. \sol{} then maps qubits to atoms based on the simulation size (1,225 qubits for the 'Atom' simulation and 256 qubits for the QuEra simulation) and a fixed 20$\times$20 AOD, initializing atom positions using the discretization scheme described earlier. \sol{} also obtains an ideal Rydberg interaction distance from Graphine.

The simulator instantiates hardware component objects adhering to real-world constraints. Atoms are initially placed in the SLM object, with SLM traps represented as an array of fixed site coordinates. Selected atoms are then "moved" into AOD traps, chosen by an algorithm iterating through all qubits in the input circuit. The AOD object contains row and column objects, which reference trapped atom objects and maintain a relative ordering constraint during movement. If AOD atoms block a move, they are recursively moved, with a hard limit of 80 recursive iterations to prevent infinite loops. Failed moves are resolved using trap changes.

During compilation, \sol{} tracks the maximum distance moved by any AOD object in each layer, which determines the time needed for all recursive moves in that layer. It also monitors failed moves to calculate necessary trap changes. After compilation, \sol{} computes the total runtime based on the number of layers, gate composition, and time required for movements and trap changes. All computed data, including layer structure, runtimes, and various statistics, is collected and returned to the user for each input circuit.

\vspace{2mm}

\noindent\textbf{Experimental Framework.} We use Python 3.11.5 for implementing \sol{}. Qiskit 0.45.0, IBM's quantum computing language~\cite{aleksandrowiczqiskit}, is used to optimize QASM circuits with its transpiler. Each quantum algorithm is read from a QASM 2.0 file into Qiskit, which contains the circuit represented in the QASM quantum assembly description language. For each input QASM file, we ran the file through the Qiskit transpiler with the highest optimization level and used the transpiler's output circuit to obtain results for \sol{} and all comparable methods. Our version of \graphine{}~\cite{patel2023graphine}, which generates the initial atom topology, uses a dual annealing optimizer derived from the SciPy 1.11.3 library~\cite{jones2016scipy} to determine atom placement. We run compilation with \sol{} and all competitors on a local research cluster with Ubuntu 22.04.2 LTS on a 32-core 2.0 GHz AMD EPYC 7551P processor with 32 GB RAM. We allow up to 24 hours for each compilation.

\vspace{2mm}

\begin{table}[t]
    \centering
    \caption{Algorithms and benchmarks used for evaluation.}
    \vspace{-1mm}
    \scalebox{0.95}{
    \begin{tabular}{ccl}
         \textbf{Acronym} & \textbf{Qubits} & \textbf{Description} \\
         \hline
         \hline
         ADD & 9 & Quantum arithmetic algorithm for adding~\cite{cuccaro2004new} \\
         ADV & 9 & Google's quantum advantage benchmark~\cite{ arute2019quantum} \\
         GCM & 13 & Generator coordinate method~\cite{li2023qasmbench} \\
         HSB & 16 & Time-dependent hamiltonian simulation~\cite{bassman2021arqtic} \\
         HLF & 10 & Hidden linear function application~\cite{bravyi2018quantum} \\
         KNN & 25 & Quantum $k$ nearest neighbors algorithm~\cite{li2023qasmbench} \\
         MLT & 10 & Quantum arithmetic algorithm for multiplying~\cite{hancock2019cirq} \\
         QAOA & 10 & Quantum alternating operator ansatz~\cite{farhi2016quantum} \\
         QEC & 17 & Quantum repetition error correction code~\cite{li2023qasmbench} \\
         QFT & 10 & Quantum Fourier transform~\cite{namias1980fractional} \\
         QGAN & 39 & Quantum generative adversarial network~\cite{li2023qasmbench} \\
         QV & 32 & IBM's quantum volume benchmark~\cite{li2023qasmbench} \\
         SAT & 11 & Quantum code for satisfiability solving~\cite{su2016quantum} \\
         SECA & 11 & Shor’s error correction algorithm~\cite{li2023qasmbench} \\
         SQRT & 18 & Quantum code for square root calculation~\cite{grover1998quantum} \\
         TFIM & 128 & Transverse-field ising model~\cite{bassman2021arqtic} \\
         VQE & 28 & Variational quantum eigensolver~\cite{li2023qasmbench} \\
         WST & 27 & W-State preparation and assessment~\cite{fleischhauer2002quantum} \\
         \hline
         \hline
    \end{tabular}}
    \vspace{-4mm}
    \label{tab:algos}
\end{table}

\noindent\textbf{Algorithms and Benchmarks.} We evaluate a variety of algorithms and benchmarks of 9-128 qubits from different application domains, covering a diverse range of circuit properties and requirements. See Table~\ref{tab:algos} for the complete list.

\vspace{2mm}

\begin{figure*}[t]
    \centering
    \includegraphics[scale=0.545]{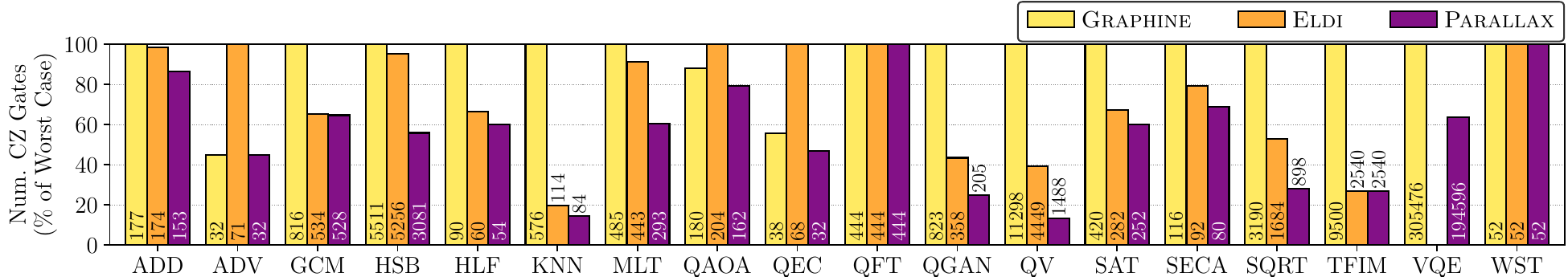}
    \vspace{1mm}
    \hrule
    \vspace{1mm}
    \caption{\sol{} achieves the fewest number of CZ gates across all algorithms on QuEra's 256-qubit computer. Note that \eldi{} did not compile the VQE algorithm within a 24-hour timeframe, and thus, the result could not be calculated. The numbers inside the bars indicate the raw CZ gate counts for different algorithms with different techniques.}
    \vspace{-6mm}
    \label{fig:czs_quera}
\end{figure*}

\begin{figure*}[t]
    \centering
    \includegraphics[scale=0.545]{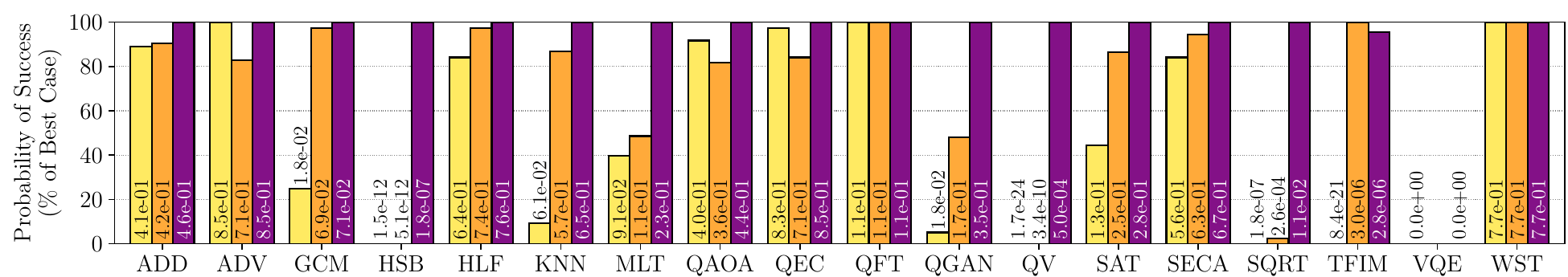}
    \vspace{1mm}
    \hrule
    \vspace{1mm}
    \caption{\sol{} achieves the highest probability of success across all algorithms on QuEra's 256-qubit computer. Note that due to its high depth and gate count, the probability of success of the VQE algorithm is too small to be calculable.}
    \vspace{-6mm}
\label{fig:success_quera}
\end{figure*}

\noindent\textbf{Competitive Techniques.} We compare \sol{} to two other neutral atom techniques: \eldi{}~\cite{Baker2021-wv,litteken2022reducing} and \graphine{}~\cite{patel2023graphine}. These techniques do not consider real hardware constraints such as the minimum trap distance and the blockade radius being $2.5\times$ the interaction radius. Thus, to make them comparable, we modified them as needed to make them hardware-compatible (e.g., we set appropriate radii and discretized them). Note that for a circuit with $q$ qubits, $g$ gates, $a$ AOD rows or columns (whichever is greater), and $s$ SLM atoms, \graphine{} has a time complexity of $O(g + q^5)$. \sol{} has a worst-case time complexity of $O(q^5 + gq^2 + a^2q^2 + ga^2s + ga^3)$.

The $q^5$ term relates to \graphine{}'s methodology for initial topology generation. The $gq^2$ term comes from computing the amount of interference caused by each CZ gate, while the $a^2q^2$ term comes from computing the number of out-of-range interactions performed for each qubit. Both $ga^2s$ and $ga^3$ are derived from calculating the worst-case number of recursive moves. Note that, like \graphine{}, the largest term for \sol{} is $O(q^5)$. Thus, both have polynomial time complexity. We could not determine the time complexity of \eldi{}; however, in practice, it had longer compilation times than \sol{} -- it could not compile for VQE (Sec.~\ref{sec:eval}). %We also execute shots of these techniques in parallel to compare with \sol{}'s parallelization.

\vspace{2mm}

\noindent\textbf{Evaluation Metrics.} We consider three metrics when comparing the techniques: 
\textbf{(1) CZ Gate Count} \textit{(lower is better)}\textbf{.} This is the number of CZ gates that each technique ends up running for a quantum algorithm. Note we do not show the U3 gate count because it remains the same for all the techniques. (2) \textbf{Probability of Success} \textit{(higher is better)}\textbf{.} This is the estimated probability of success of a circuit after taking a product of the error rates of all circuit components and including decoherence error~\cite{tannu2019ensemble,patel2020veritas,patel2023graphine}. (3) \textbf{Circuit Runtime} \textit{(lower is better)}\textbf{.} This is the time to run one logical shot of a circuit. (4) \textbf{Total Execution Time} \textit{(lower is better)}\textbf{.} This is the total time to run 8,000 logical shots per circuit to generate the output probability distribution. This time is affected by circuit runtime and how many logical shots are parallelized.
\section{Evaluation}
\label{sec:eval}

%\begin{figure*}[t]
%    \centering
%    \includegraphics[scale=0.545]{figures/quera_figs/depths.pdf}
%    \vspace{1mm}
%    \hrule
%    \vspace{1mm}
%    \caption{\sol{} can have a higher circuit runtime than competitive techniques on QuEra's 256-qubit computer. See Fig. \ref{fig:czs_quera} for legend.}
%    \vspace{-6mm}
%\label{fig:depths_quera}
%\end{figure*}

%\begin{figure*}[t]
%\centering
%\includegraphics[scale=0.545]{figures/atom_figs/depths.pdf}
%    \vspace{1mm}
%    \hrule
%    \vspace{1mm}
%    \caption{\sol{}'s circuit runtime differential diminishes as we scale to Atom's 1,225-qubit quantum computer. See Fig. \ref{fig:czs_quera} for legend.}
%    \vspace{-6mm}
%\label{fig:depths_atom}
%\end{figure*}

% by 34\% compared to \graphine{} and 28\% compared to \eldi{} using Atom's 1,225-atom system.
\begin{table*}[t]
    \centering
    \caption{\sol{} can have a higher circuit runtime (in $\mu$s) than competitive techniques on QuEra's 256-qubit computer. However, this runtime differential diminishes considerably as we scale to Atom's 1,225-qubit quantum computer.}
    \vspace{-1mm}
    \scalebox{0.88}{
    \begin{tabular}{c|cc|cc|cc|cc|cc|cc|cc|cc|cc}
          & \multicolumn{2}{c|}{\textbf{ADD}} & \multicolumn{2}{c|}{\textbf{ADV}} & \multicolumn{2}{c|}{\textbf{GCM}} & \multicolumn{2}{c|}{\textbf{HSB}} & \multicolumn{2}{c|}{\textbf{HLF}} & \multicolumn{2}{c|}{\textbf{KNN}} & \multicolumn{2}{c|}{\textbf{MLT}} & \multicolumn{2}{c|}{\textbf{QAOA}} & \multicolumn{2}{c}{\textbf{QEC}} \\
         \hline
         \textbf{Num. Qubits} & \textbf{256} & \textbf{1,225} & \textbf{256} & \textbf{1,225} & \textbf{256} & \textbf{1,225} & \textbf{256} & \textbf{1,225} & \textbf{256} & \textbf{1,225} & \textbf{256} & \textbf{1,225} & \textbf{256} & \textbf{1,225} & \textbf{256} & \textbf{1,225} & \textbf{256} & \textbf{1,225} \\
         \hline
         \textbf{\eldi{}} & 372 & 386 & 99 & 112 & 1460 & 1460 & 7942 & 8299 & 116 & 130 & 203 & 202 & 769 & 807 & 389 & 366 & 96 & 98 \\
         \textbf{\graphine{}} & 386 & 425 & 68 & 68 & 1685 & 1830 & 8140 & 8142 & 142 & 117 & 584 & 374 & 804 & 810 & 368 & 355 & 74 & 68 \\
         \textbf{\sol{}} & 371 & 409 & 67 & 67 & 1530 & 1705 & 7180 & 8431 & 125 & 130 & 3239 & 282 & 700 & 764 & 357 & 352 & 69 & 67 \\
         
         \multicolumn{19}{c}{} \\
         & \multicolumn{2}{c|}{\textbf{QFT}} & \multicolumn{2}{c|}{\textbf{QGAN}} & \multicolumn{2}{c|}{\textbf{QV}} & \multicolumn{2}{c|}{\textbf{SAT}} & \multicolumn{2}{c|}{\textbf{SECA}} & \multicolumn{2}{c|}{\textbf{SQRT}} & \multicolumn{2}{c|}{\textbf{TFIM}} & \multicolumn{2}{c|}{\textbf{VQE}} & \multicolumn{2}{c}{\textbf{WST}} \\
         \hline
         \textbf{Num. Qubits} & \textbf{256} & \textbf{1,225} & \textbf{256} & \textbf{1,225} & \textbf{256} & \textbf{1,225} & \textbf{256} & \textbf{1,225} & \textbf{256} & \textbf{1,225} & \textbf{256} & \textbf{1,225} & \textbf{256} & \textbf{1,225} & \textbf{256} & \textbf{1,225} & \textbf{256} & \textbf{1,225} \\
         \hline
         \textbf{\eldi{}} & 966 & 972 & 531 & 469 & 5240 & 5161 & 633 & 632 & 180 & 180 & 2561 & 2558 & 1755 & 1319 & N/A & N/A & 108 & 108 \\
         \textbf{\graphine{}} & 972 & 1055 & 927 & 772 & 1.1$e$4 & 9288 & 749 & 748 & 199 & 218 & 3841 & 2250 & 7365 & 4487 & 6.1$e$5 & 5.9$e$5 & 108 & 108 \\
         \textbf{\sol{}} & 971 & 1016 & 3342 & 1841 & 5.7$e$4 & 2.9$e$4 & 663 & 727 & 188 & 219 & 1.9$e$4 & 2199 & 5.3$e$4 & 11153 & 6.5$e$5 & 7.2$e$5 & 108 & 108 \\
    \end{tabular}}
    \vspace{-4mm}
    \label{tab:runtimes}
\end{table*}

\begin{figure*}[t]
    \centering
    \includegraphics[scale=0.425]{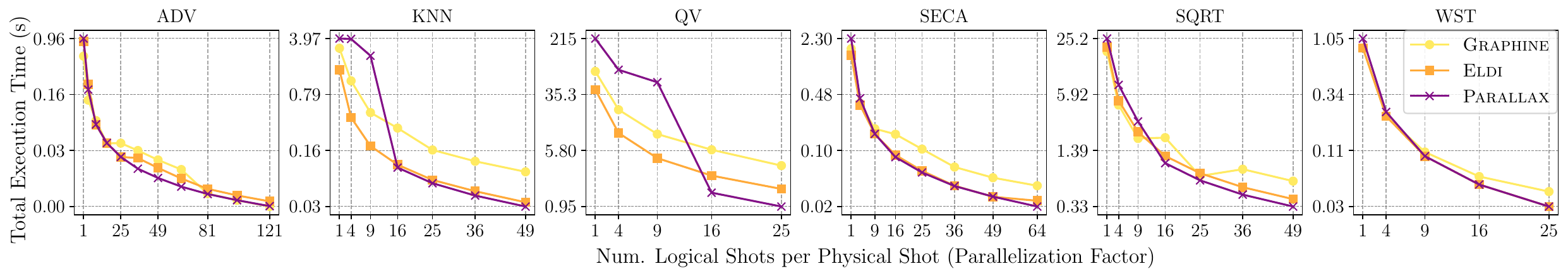}
    \vspace{0.5mm}
    \hrule
    \vspace{1mm}
    \caption{\sol{} parallelizes the logical shots of the circuits being run on one computer (Atom's 1,225-qubit computer in this case): the total execution time decreases. The figure also shows parallelized \eldi{} and \graphine{}. Note the log scales.}
    \vspace{-6mm}
\label{fig:parallelism}
\end{figure*}

\noindent\textbf{On average, \sol{} reduces the number of two-qubit CZ gates required for circuit execution by 39\% compared to \graphine{} and 25\% compared to \eldi{} on a 256-atom system}. As shown in Fig.~\ref{fig:czs_quera}, \sol{} has the fewest CZ gate counts for all the algorithms. In most cases, \graphine{} has the highest counts, while in a few of the cases, \eldi{} does. \sol{} has the fewest CZ gate counts due to the fact that it has no SWAP operations, while \graphine{} and \eldi{} do (each SWAP is composed of three CZ gates). Because of this, \sol{} has at most the same CZ count as the other two methods and, in the majority of cases, significantly fewer. \sol{}'s reduction in CZ gate counts is larger for algorithms that have higher qubit connectivity, where connectivity refers to how many other qubits each qubit interacts with via CZ gates. The reason for this is that higher connectivity implies increased difficulty in placing interacting qubits close together since more atoms need to be within the Rydberg radius of each other. This can be seen in the comparison between the TFIM and QV algorithms. 

TFIM is a structured algorithm where each qubit interacts with at most two other qubits, meaning it is a good example of a low-connectivity algorithm. \sol{} does not improve on CZ gate counts when compared to \eldi{} in this case since atoms can mostly execute CZ gates without the need for any SWAPs/moves. This is in contrast to QV, which has a more diverse structure and higher connectivity between qubits. With QV, \sol{} has a 67\% reduction in CZ gate counts compared to \eldi{} and an 87\% reduction compared to \graphine{}. \sol{} similarly improves the CZ gate counts of other algorithms by eliminating SWAPs.

Note also that we do not show the U3 gate counts for brevity as they are the same across all techniques, as the techniques only deal with CZ gates due to their much higher error rates.

\vspace{2mm}

\noindent\textbf{\sol{} thus improves the probability of success for a noisy quantum circuit execution by 46\% compared to \graphine{} and 28\% compared to \eldi{}, on average.} As shown in Fig.~\ref{fig:success_quera}, \sol{} achieves a higher probability of success than other techniques across algorithms (only TFIM is slightly lower). This result is primarily a function of a reduction in the number of CZ gates that \sol{} achieves.

While this is a positive result, as can be seen in Table~\ref{tab:runtimes}, \sol{} has a longer runtime for some algorithms on the 256-qubit computer. Note that this is the runtime of a single circuit, not the total execution time of all the shots. Note also that this increase in runtime does not lead to a lower probability of success (as we saw in Fig.~\ref{fig:success_quera}) because of the large coherence times of neutral atom qubits. The increase in runtime is not necessarily because of atom movement. Each SWAP takes about 2.4$\mu$s, whereas atoms move at a rate of 55$\mu$m$/\mu$s. This means that on the 256-atom system, the longest possible move would take about 2$\mu$s, which is still lower than the time it takes to SWAP.

The cause of the increase in runtime is the trap changes required to interact with the static atoms in the topology. Whenever AOD movement fails, which is mainly due to two immobile SLM atoms being out of Rydberg range for a CZ gate, \sol{} must execute a time-consuming AOD trap-change operation. It switches the traps of one of the atoms into a new AOD location and then moves it into the Rydberg range, ending by returning it to its original SLM trap. Each trap-change operation takes roughly 100$\mu$s~\cite{regional_addressing}, which is significantly longer than the 2.4$\mu$s required for a SWAP gate used by techniques like \eldi{} and \graphine{}. However, it's important to note that while trap changes are time-consuming, they have a low error rate (less than 0.1\%~\cite{regional_addressing}), compared to the much higher error rate of SWAP gates (1.43\% per SWAP~\cite{highfidbluv}). This trade-off between runtime and error rate contributes to \sol{}'s higher overall success rates.

The number of times these trap changes occur depends on the initial topology. If the circuit is given too small a physical space, then \sol{} is unable to create a good initial topology optimally and instead places atoms wherever there is free space rather than their ideal locations, given the discretization constraints. Thus, \sol{} performs better in terms of circuit runtimes as computer sizes scale up -- compare the results for the 256-qubit computer vs. the 1,225-qubit computer in Table~\ref{tab:runtimes}. This can be seen with TFIM; mapping the 128 logical qubits of TFIM to the 256-atom QuEra system results in a sub-optimal initial topology since there simply isn't enough space to place the atoms optimally. This also explains why TFIM is the only circuit where \sol{} has a slightly lower success rate than \eldi{}. However, when \sol{} has more space to work with in the 1,225-atom computer, the initial topology is much closer to the optimal layout. Thus, the runtime of TFIM decreases on the 1,225-qubit computer.
Similarly, in Table~\ref{tab:runtimes}, there are multiple circuits where \sol{} runs faster than competing techniques for the 1,225-qubit computer. As a note, CZ gate counts and probability of success remain unaffected by the size of the computer; therefore, we do not repeat the results. 

It's worth noting that while \sol{} may incur longer compilation times on classical computers for large circuits, these classical overheads do not add to the quantum error rates. Moreover, \sol{} does not introduce additional quantum overheads in terms of increased qubits, longer depths, or higher quantum costs. Despite occasional longer runtimes, \sol{}'s higher success rates generally justify the increased classical compilation times and occasional longer quantum runtimes. In fact, \sol{} further reduces runtime by utilizing parallelism, as discussed next.

\vspace{2mm}

\begin{figure*}[t]
    \centering
    \hspace{5mm}
    \includegraphics[scale=0.555]{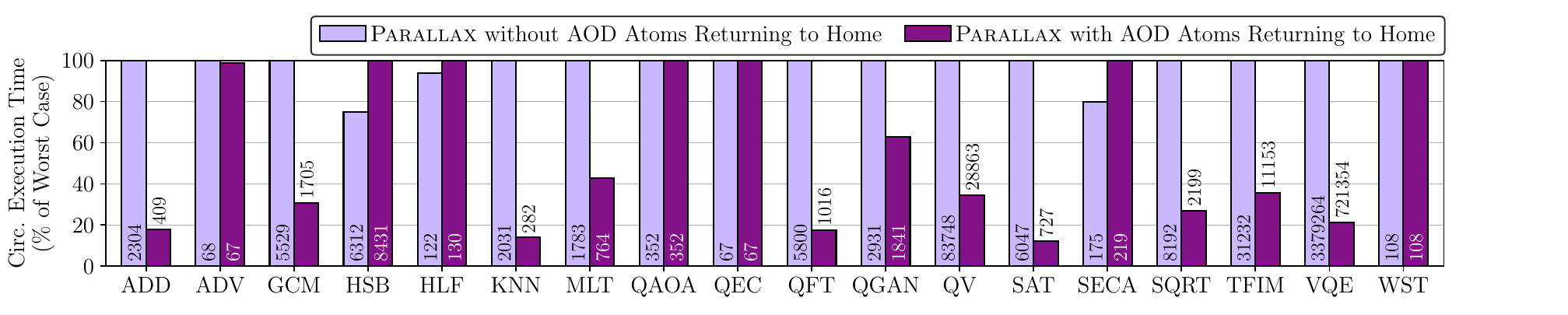}
    \vspace{-4.5mm}
    \hrule
    \vspace{1mm}
    \caption{Runtime comparison between having AOD qubits return to their initial location after having moved vs. not returning.}
    \vspace{-6mm}
    \label{fig:home_runtime}
\end{figure*}
\begin{figure*}[t]
    \centering
    \hspace{3.5mm}
    \includegraphics[scale=0.555]{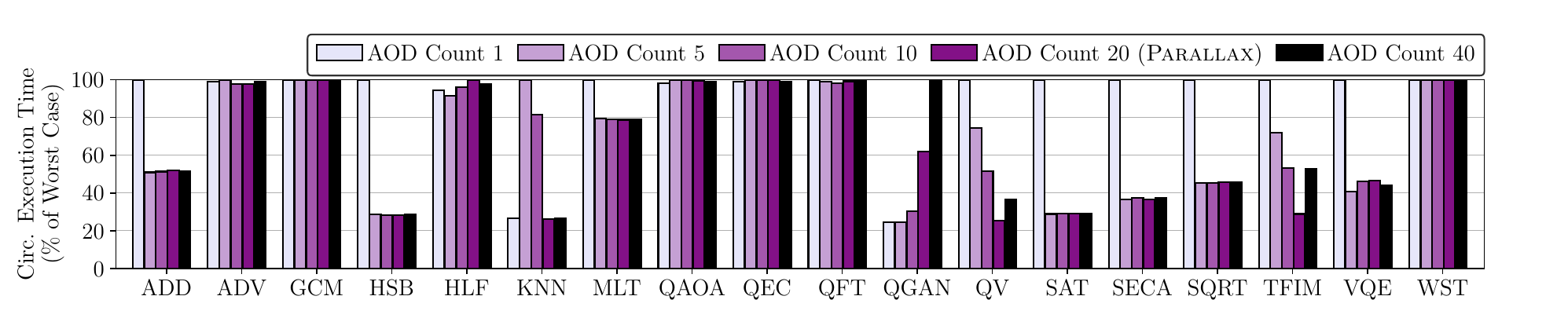}
    \vspace{-4.5mm}
    \hrule
    \vspace{1mm}
    \caption{\sol{}'s runtimes with varying AOD row and column counts. ``AOD Count 20'' means 20 AOD rows and columns.}
    \vspace{-6mm}
    \label{fig:aods_time}
\end{figure*}

\noindent\textbf{\sol{}'s parallelism-centered design reduces the total execution time by 97\% on average compared to it running one shot at a time.} Note that we obtained simulated results only for the larger Atom computer to showcase the effects of parallelism better. Due to the innate parallelizable design of \sol{}, it can run as many logical shots of the same circuit in a single physical shot as there are free atoms, space, and AOD rows/columns in the hardware. As an example, for a relatively small circuit like ADV with nine qubits, \sol{} can run as many as 121 copies of the circuit in parallel on the 1,225-atom computer, shown in Fig.\ref{fig:parallelism} (the figure only shows a few examples for brevity). For this example, each AOD row and column stores 11 atoms given an AOD size of 20 AOD rows and columns~\cite{adamsaod24}. This demonstrates \sol{}'s ability to efficiently utilize all available resources of a neutral atom computer for small circuits by parallelizing across many copies. Conversely, it has demonstrated the ability to compile even complex circuits like the 450,000-gate VQE circuit, whose execution is beyond the capabilities of current computers. This dual approach to scalability—parallelizing small circuits and efficiently handling large ones—positions \sol{} to provide effective compilation solutions as quantum hardware capabilities evolve.

For the sake of comparability, we also parallelized \eldi{} and \graphine{} to execute multiple copies of compiled circuits using the two techniques on the same 1,225-atom computer. Both of these techniques saw similar execution time speedups, with \sol{} and \eldi{} performing comparably. In some cases, such as QV, while \sol{} has a worse total execution time than \eldi{} and \graphine{} without parallelism, it performs better with higher parallelism. This highlights the novelty of \sol{}, which is being able to execute parallel circuits while utilizing AOD movements, which is challenging compared to parallelizing circuits on stationary qubits with \eldi{} and \graphine{}.

Next, we perform ablation studies for \sol{}.

\vspace{2mm}

\noindent\textbf{\sol{} has a 40\% lower circuit runtime and an 8.9\% higher success rate when having AOD qubits return to their initial home locations after being moved to execute a CZ gate compared to not returning them.} As mentioned previously, all atoms are initialized to optimal locations by \graphine{}, where they are closer to atoms they interact with frequently. By returning the AOD atoms to be near their optimal initial locations after movement (e.g., returning them home), we reduce the amount of AOD movement needed for future CZ gates and thus reduce the circuit runtime. See Fig.~\ref{fig:home_runtime} for results on circuit runtimes between AOD atoms returning to their home locations as compared to not returning. Note: the above ablation has no impact on the CZ gate count, and thus, the probability of success is also negligibly affected.

\vspace{2mm}

\noindent\textbf{When comparing different AOD row and column counts, \sol{}'s usage of 20 AOD rows and columns has the lowest circuit execution time of all tested AOD counts. On average, the 20-count variant had 36\% lower runtime than the worst case for each algorithm (see Fig.~\ref{fig:aods_time}).} The 1-count variant achieves a 9\% lower average runtime than the worst case across all algorithms, the 5-count has 29\% lower, the 10-count has 32\% lower, and the 40-count has 32\% lower. Note: the above ablation has no impact on the CZ gate count, and thus, the probability of success is also negligibly affected.

\section{Discussion and Conclusion}

We proposed \sol{} as a novel compiler for neutral atom systems. By eliminating suboptimal SWAP operations, balancing qubit mobility, and introducing parallelization with atom movements, \sol{} significantly enhances the efficiency and performance of neutral atom systems. We believe this will serve as an important step toward realizing the full potential of quantum computing with neutral atoms.

\sol{} is designed with an eye toward the future of neutral atom quantum computing. While individually addressed Rydberg excitations have been demonstrated experimentally, \sol{} anticipates the scaling up of these capabilities to larger systems. Our open-source simulator allows for easy updates to technology parameters like AOD count and atom movement speed, ensuring \sol{} can evolve alongside advancements in neutral atom hardware.

\section*{Acknowledgement}

We would like to thank the anonymous reviewers for their feedback. This work was supported by Rice University. 

\balance
\bibliographystyle{IEEEtranS}
\bibliography{main}

\end{document}